\documentclass[pra, aps, showpacs, twocolumn]{revtex4}

\usepackage{amsmath}
\usepackage{mathrsfs}
\usepackage[dvips]{graphicx}
\usepackage{subfigure}
\usepackage{array}
\usepackage{amssymb}
\usepackage{mathbbol}

\DeclareGraphicsExtensions{.eps}

\newcommand{\ket}[1]{| #1\rangle}
\newcommand{\bra}[1]{\langle #1 |}
\newcommand{\bracket}[2]{\langle #1 | #2 \rangle}
\def\tr{\mathrm{tr}}

\begin{document}
\title{Multiqubit symmetric states with high geometric entanglement}
\date{\today}

\author{J. Martin,$^1$ O. Giraud,$^{2,3,4,5}$ P. A.~Braun,$^{6,7}$ D. Braun,$^{2,3}$ and T. Bastin$^1$}
\address{$^{1}$Institut de Physique Nucl\'eaire, Atomique et de Spectroscopie, Universit\'e de Li\`ege, 4000 Li\`ege, Belgium\\
$^{2}$Universit\'e de Toulouse; UPS; Laboratoire de Physique Th\'eorique (IRSAMC); 31062 Toulouse, France\\
$^{3}$CNRS; LPT (IRSAMC); 31062 Toulouse, France\\
$^{4}$Universit\'e Paris-Sud, LPTMS, UMR8626, B\^at. 100, Universit\'e Paris-Sud, 91405 Orsay, France\\
$^{5}$CNRS,  LPTMS, UMR8626, B\^at. 100, Universit\'e Paris-Sud, 91405 Orsay, France\\
$^{6}$Fachbereich Physik, Universit\"at Duisburg--Essen, 47048 Duisburg, Germany\\
$^{7}$Institute of Physics, Saint-Petersburg University, 198504 Saint-Petersburg,  Russia}

\begin{abstract}
We propose a detailed study of the geometric entanglement properties of pure symmetric $N$-qubit states, focusing more particularly on the identification of symmetric states with a high geometric entanglement and how their entanglement behaves asymptotically for large $N$. We show that much higher geometric entanglement with improved asymptotical behavior can be obtained in comparison with the highly entangled balanced Dicke states studied previously. We also derive an upper bound for the geometric measure of entanglement of symmetric states. The connection with the quantumness of a state is discussed.
\end{abstract}


\maketitle

\section{Introduction}
\label{Introduction}

Entanglement of compound systems is known to be a key resource in quantum information allowing non-local correlations~\cite{Nie00}. In many practical applications it is of fundamental importance to know whether a state is entangled or not. Very often, however, this information is not sufficient and it is also required to know \emph{how much} a state is entangled. A useful tool to quantify the amount of entanglement of a state is given by the so-called entanglement measures~\cite{Ple07}. A widely used such measure is provided by the geometric measure of entanglement that is defined for a pure state $|\psi\rangle$ by~\cite{Wei03}
\begin{equation}\label{geoent}
E_G(\ket{\psi})= 1-\max_{\ket{\Phi} = \ket{\phi_1,\phi_2,\phi_3,\ldots}} |\bracket{\psi}{\Phi}|^2
\end{equation}
where the maximum is taken over all separable pure states. The extension to mixed states is done via the usual convex-roof construction~\cite{Ple07}. The geometric measure fulfills all the requirements of an entanglement monotone~\cite{Wei03} and is comprised between 0 and 1. While the lower value is realized for separable states, the upper value is never reached because that would imply having a state orthogonal to all separable states and this is simply impossible. The geometric measure of entanglement has been shown to have several operational interpretations, on one side in terms of the utility of a state as an initial state for the Grover algorithm~\cite{Bih02}, and on the other side in terms of the difficulty of state discrimination under local operations and classical communications (LOCC)~\cite{Hay06}. It is also related to optimal entanglement witnesses~\cite{Wei03, Hay08} and has nice applications in many-body physics and condensed matter systems~\cite{Wei05, Mar08, Oru08, Nak09}.

Despite its usefulness, the explicit value of the geometric measure of entanglement has only been derived, so far, for a limited number of entangled states, such as $N$-qubit Greenberger-Horne-Zeilinger (GHZ) states~\cite{Wei03}, Dicke states~\cite{Wei03}, generalized $W$-states~\cite{Tam10}, graph states~\cite{Mar07}, or other typical states with given symmetry properties~\cite{Hay08, Tam08, Mar09, Par09, Che09}. The geometric measure remains unknown for most of the multipartite states simply because of the definition that involves an optimization procedure over the class of separable states and this represents a formidable task in the general case even with numerical approaches. A very recent study showed, however, that this task is drastically simplified in the case of states that are symmetric under any permutation of the parties~\cite{Hub09}. In that case the optimization can be done on the restricted set of symmetric separable states $|\phi, \ldots, \phi\rangle$. This represents a huge simplification in the calculation of the geometric measure since the number of parameters involved in the optimization procedure does not depend on the number of parties anymore. For qubit systems, only two real parameters are required. In this paper we exploit this great simplification to propose a detailed study of the geometric entanglement properties of symmetric $N$-qubit states. We focus more particularly on the identification of symmetric states with a high geometric entanglement measure and how their entanglement behaves asymptotically for large $N$. We show that much higher geometric entanglement with improved asymptotical behavior can be obtained in comparison with the highly entangled balanced Dicke states~\cite{Wei03}. We also derive an upper bound for the geometric measure of entanglement of symmetric states, that is significantly tighter than the one known to hold for any state~\cite{Jun08}. We finally make a connection between the most geometrically entangled states and states having the highest quantumness as defined in Ref.~\cite{Q5}.

The paper is organized as follows. In Sec.~\ref{SecGMEdef} the useful representations of $N$-qubit symmetric states are given to get a simplified expression of the geometric measure of entanglement for that class of states. In Sec.~\ref{SecUB} we derive the announced upper bound. In Sec.~\ref{SecHighGME} we investigate various configurations of highly geometrically entangled symmetric states. Their asymptotic behavior with respect to the number $N$ of qubits is analyzed. The connection of the geometric entanglement with the quantumness of a state is given in Sec.~\ref{SecQuantumness}. We finally draw conclusion in Sec.~\ref{SecConclusion}.

\section{Geometric entanglement for symmetric states}
\label{SecGMEdef}

In an $N$-qubit system, an arbitrary symmetric state is conveniently represented in either the Majorana~\cite{Maj32} or the Dicke state~\cite{Dic54} representation.
In the Majorana representation, any symmetric state $|\psi_S\rangle$ is univocally specified by $N$ single-qubit states $|\phi_i\rangle \equiv \alpha_i |0\rangle + \beta_i |1\rangle$ ($i=1,\ldots, N$) through a sum over all permutations $\sigma$ of the qubits~:
\begin{equation}
\label{psisym}
|\psi_S\rangle = \mathcal{N} \sum_{\sigma} \ket{\phi_{\sigma(1)},\ldots,\phi_{\sigma(N)}},
\end{equation}
where $\ket{\phi_{\sigma(1)},\ldots,\phi_{\sigma(N)}}$ denotes the product state $\ket{\phi_{\sigma(1)}}\otimes\ldots\otimes\ket{\phi_{\sigma(N)}}$ and $\mathcal{N}$ is a normalization prefactor. In the Bloch sphere picture~\cite{Ben06}, the single-qubit states $\ket{\phi_{i}}$ can be represented as points on the unit sphere labelled by two angles $(\theta_i,\varphi_i)$ with $\alpha_i=\cos\left(\theta_i/2\right)$ and $\beta_i=e^{i\varphi_i}\sin\left(\theta_i/2\right)$. Any arrangement of $N$ points on the unit sphere thus defines univocally a symmetric state in the Majorana representation.

A particularly simple yet important example is given by the symmetric Dicke states with $k$ excitations
\begin{equation}
\label{Dicke}
\ket{D_N(k)} = \frac{1}{\sqrt{C_N^k}} \sum_{\sigma} \ket{\underbrace{0\ldots 0}_{N-k}\underbrace{1\ldots 1}_{k}},
\end{equation}
with $C_N^k$ the binomial coefficient of $N$ and $k$. In the Majorana representation, the Dicke states correspond to $N-k$ points at the North pole ($\theta = 0$) and $k$ points at the South pole of the Bloch sphere.

In the Dicke state representation, any symmetric state $|\psi_S\rangle$ is merely expanded in the orthonormal basis formed by the $N+1$ Dicke states in the symmetric subspace~:
\begin{equation}
\label{psisymD}
|\psi_S\rangle = \mathcal{N}\,\sum_{k=0}^N d_k |D_N(k) \rangle,
\end{equation}
with $d_k$ ($k=0,\ldots,N$) the complex expansion coefficients.

The Dicke state representation of a symmetric state written in the form of Eq.~(\ref{psisym}) is obtained through the relation~\cite{Bas09a}
\begin{equation}
\label{dk} d_k = \sqrt{C_N^k} \sum_{\sigma}
    \beta_{\sigma(1)} \ldots \beta_{\sigma(k)} \alpha_{\sigma(k+1)} \ldots
    \alpha_{\sigma(N)}.
\end{equation}
Inversely, the Majorana representation of a symmetric state expressed in the Dicke state basis is obtained with $N$ single-qubit $|\phi_i\rangle$ states defined by $\alpha_i/\beta_i$ equal to the $K$ roots of the polynomial $P(z) = \sum_k^N (-1)^{k} (C_N^k)^{1/2} d_k z^k$, $K$ being the polynomial degree and the remaining $\alpha_i$ equal to 1~\cite{Bas09b}.

The Majorana representation is particularly useful to get a simplified expression of the geometric measure of entanglement for any symmetric $N$-qubit state $|\psi_S\rangle$. Inserting Eq.~(\ref{psisym}) into Eq.~(\ref{geoent}) and considering that the maximization is only required over the set of symmetric separable states $|\phi, \ldots, \phi\rangle$~\cite{Hub09} yields immediately
\begin{equation}\label{geoentsym}
E_G(\ket{\psi_S}) = 1-\mathcal{N}^2 N!^2 \max_{\ket{\phi}}\prod_{i=1}^N |\bracket{\phi_i}{\phi}|^2.
\end{equation}

The geometric measure of entanglement of the $N$-qubit GHZ states $|\textrm{GHZ}_N\rangle = (|0\ldots0\rangle + |1\ldots1\rangle)/\sqrt{2}$ is equal to $1/2$ regardless the number of qubits~\cite{Wei03}. For the symmetric Dicke states, it reads~\cite{Wei03}
\begin{equation}\label{EGDicke}
E_G(\ket{D_N(k)}) =1-C_N^k\left(\frac{k}{N}\right)^k\left(\frac{N-k}{N}\right)^{N-k}.
\end{equation}
It is maximal for a balanced number of excitations, i.e., for $k = k_N$ with $k_N$ closest to $N/2$.
In that case we get the asymptotic behavior for large $N$
\begin{equation}\label{entDicke}
 E_G(\ket{D_N(k_N)})= 1-\sqrt{\frac{2}{\pi N}}+\mathcal{O}(N^{-3/2}).
\end{equation}
If the right-hand side of Eq.~(\ref{entDicke}) converges to 1 with the number of qubits, this convergence can be qualified as quite slow since the difference with one only decreases as the inverse of the square root of $N$. In the next section, we derive an upper bound of the geometric measure of entanglement that allows one to expect symmetric states with a geometric entanglement converging much faster to one with the number of qubits. Various configurations of symmetric states are identified with that behavior.

\section{Geometric entanglement upper bound}
\label{SecUB}

The geometric measure of entanglement of an $N$-qubit symmetric state $|\psi_S\rangle$ necessarily verifies
\begin{equation}\label{upperbound}
E_G(\ket{\psi_S}) < 1-\frac{1}{N+1}.
\end{equation}

Equation~(\ref{upperbound}) is immediately obtained from the representation of the identity in the $(N+1)$-dimensional subspace of symmetric states as a combination of all projectors onto the symmetric separable states $\ket{\Phi}=\ket{\phi,\ldots, \phi}$~\cite{Bar97}~:
\begin{equation}\label{id}
\mathbb{1}_{N+1}=\frac{N+1}{4\pi}\int_0^\pi\sin\theta d\theta\int_0^{2\pi}d\varphi\,\ket{\Phi}\bra{\Phi},
\end{equation}
where the single-qubit state $\ket{\phi}$ is parametrised by the two angles ($\theta$,$\varphi$) according to $|\phi\rangle=\cos\left(\theta/2\right)\ket{0}+e^{i\varphi}\sin\left(\theta/2\right)\ket{1}$. Since obviously the maximal overlap of $\ket{\psi_S}$ with a symmetric separable state is always larger than the average overlap over all symmetric separable states $\ket{\Phi}$, we have
\begin{equation}
\begin{aligned}
\max_{\ket{\Phi}}|\bracket{\psi_S}{\Phi}|^2 & > \frac{1}{4\pi}\int_0^\pi\sin\theta d\theta\int_0^{2\pi}d\varphi\,|\bracket{\psi_S}{\Phi}|^2\\
& >\big\langle\psi_S\big|\frac{\mathbb{1}_{N+1}}{N+1}\big|\psi_S\big\rangle\\
& > \frac{1}{N+1},
\end{aligned}
\end{equation}
which proves the upper bound (\ref{upperbound}). It is interesting to note that this upper bound is significantly tighter than the one known to hold for any state $|\psi\rangle$ beyond the symmetric subspace given by~\cite{Jun08}
\begin{equation}
E_G(\ket{\psi}) < 1-\frac{1}{2^{N-1}}.
\end{equation}

\section{High geometric entanglement configurations}
\label{SecHighGME}

\subsection{Highest geometric entanglement configurations}
\label{Highest}

\begin{figure}
\begin{center}
\includegraphics[clip=true,width=0.85\linewidth]{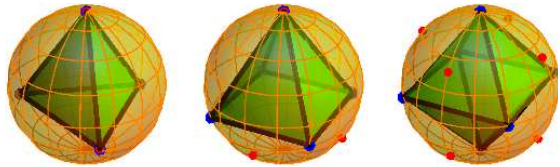}
\end{center}
\caption{Majorana representation of the maximally entangled symmetric states for $N=4-6$ (from left to right). The blue dots are the Majorana representation points of the maximally entangled symmetric states. The red dots correspond to all closest separable states. There are as many of them as the number of polyhedron faces.}
\label{polyhedraFig}
\end{figure}

Although the calculation of the geometric measure of entanglement is greatly simplified for the particular case of $N$-qubit symmetric states, the quest of the states in the symmetric subspace having the highest geometric entanglement remains a task that cannot be solved analytically in the general $N$ case. Even numerically the task gets very quickly extremely involved for increasing $N$ values. Here we report results up to $N=6$~\cite{Mar10}. For $N = 2$, the geometric measure of entanglement of a state is equal to the minimal Schmidt coefficient~\cite{Wei03} and the most entangled symmetric state is thus given by the Bell state $(|00\rangle + |11\rangle)/\sqrt{2}$ with $E_G = 1/2$. For $N = 3$, it is given by the $W$ state $|D_3(1)\rangle$ with $E_G = 5/9$~\cite{Che09}. For $N = 4$ to $6$, the Bloch sphere Majorana representation points of the most entangled symmetric states we identified numerically are reported in Table~\ref{tab} and further illustrated in Fig~\ref{polyhedraFig}. It can be noticed from the figure that these states are characterized by $N$ distinct points on the Bloch sphere with a large spread, similarly as $N$ equal electrical charges tend to place as far as possible from each other when they are constrained to a conducting sphere (Thomson problem). For $N = 4$ and $6$, the points are vertices of two platonic regular polyhedra, namely the tetrahedron and the octahedron, respectively. For $N=5$, the point configuration is a square pyramid. The Dicke state representations of these most entangled symmetric states read $\sqrt{1/3}(\ket{D_4(0)}+\sqrt{2}\,\ket{D_4(3)})$ for $N=4$, $\sqrt{1-\xi^2}\,\ket{D_5(0)}-\xi\,\ket{D_5(4)}$ with $\xi=0.8373\ldots$ for $N=5$, and $\sqrt{1/2}\,(\ket{D_6(1)}+\ket{D_6(5)})$ for $N=6$.

\begin{table}
\begin{center}
\begin{tabular}{|c|p{160pt}|c|}
\hline
$N$ & \centering $(\theta_i,\varphi_i)$ & $E_G$\\ \hline\hline
4 & $(0,0)$, $(\theta_0,0)$, $(\theta_0,2\pi/3)$, $(\theta_0,4\pi/3)$ with $\theta_0=\mathrm{arccos}(-1/3)$ & 2/3=0.6666\ldots\\ \hline
5 & $(0,0)$, $(\theta_0,0)$, $(\theta_0,\pi/2)$, $(\theta_0,\pi)$, $(\theta_0,3\pi/2)$ with $\theta_0=1.8737\ldots$ & 0.7011\ldots\\ \hline
6 & $(0,0)$, $(\pi,0)$, $(\pi/2,0)$, $(\pi/2,\pi/2)$, $(\pi/2,\pi)$, $(\pi/2,3\pi/2)$ & 7/9=0.7777\ldots\\ \hline
\end{tabular}
\end{center}
\caption{Most entangled symmetric states defined by their set of angles ($\theta_i, \varphi_i$). 
For $N=5$, $\cos\theta_0\equiv  x$ is given by the largest real root of the polynomial  $101 + 362 x + 308 x^2 + 894 x^3 + 670 x^4 + 894 x^5 + 308 x^6 + 362 x^7 + 101 x^8$, and the corresponding geometric entanglement $E_G\equiv y$ is given by the smallest real root of the polynomial $25 - 475 y + 7630 y^2 - 18980 y^3 + 12824 y^4$.}
\label{tab}
\end{table}

\subsection{Coulomb and related arrangements}

\begin{figure}
\begin{center}
\includegraphics[width=.95\linewidth]{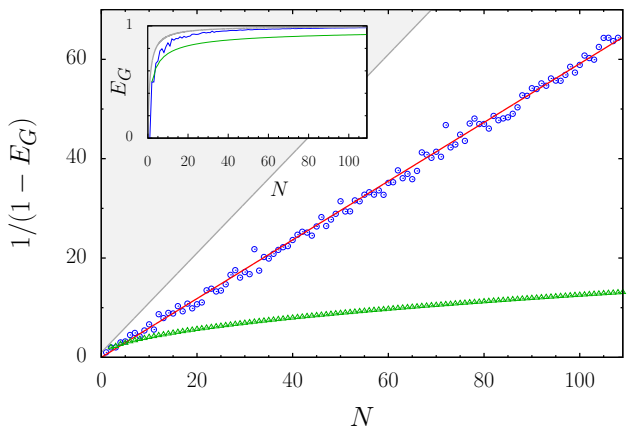}
\end{center}
\caption{Geometric entanglement of symmetric states for the Coulomb arrangement for $N$ up to $110$ (blue circles).
Equation~(\ref{entCoul}) is displayed in red. Green triangles correspond to the entanglement of symmetric Dicke states with a balanced number of excitations given by Eq.~(\ref{entDicke}). The grey shaded area shows the domain ruled out by the upper bound (\ref{upperbound}).}
\label{figCoul}
\end{figure}

Although similar, the Thomson problem mentioned above remains distinct from the quest of maximal entanglement configurations since it requires to find $N$ charge positions $\mathbf{r}_i$ on a sphere with the different constraint of minimizing the total electrostatic energy
\begin{equation}\label{enCoul}
E=\sum_{i=1}^N\sum_{j>i}^N\frac{1}{|\mathbf{r}_i-\mathbf{r}_j|}.
\end{equation}
These configurations of points, which we call the Coulomb arrangements, are given in Ref.~\cite{Sloane} for $N$ up to 130. For $N = 4$ to $6$, they coincide with the Majorana representation points of the most geometrically entangled symmetric states, except for $N = 5$ where the Coulomb arrangement is a triangular dipyramid corresponding to a geometric entanglement of 0.6875. The Coulomb arrangement is thus proven not to provide the best point configuration for the highest geometric entanglement of symmetric states. It is nevertheless expected to provide high entanglement configurations in view of the large spread it leads to. We illustrate in Fig.~\ref{figCoul} the geometric entanglement of the symmetric states with Majorana representation points distributed according to the Coulomb arrangements. These states are denoted by $\ket{\psi_{\mathrm{Coul}}(N)}$. We find numerically that the entanglement of these states scales up to small fluctuations like
\begin{equation}\label{entCoul}
E_G(\ket{\psi_{\mathrm{Coul}}(N)})\simeq 1-\frac{C}{N+1},
\end{equation}
with $C \approx 1.71$ a numerical constant. It is very interesting to note that the Coulomb arrangements, though they do not necessarily provide the most geometrically entangled symmetric states, define nevertheless states with a very high geometric entanglement close to the upper bound~(\ref{upperbound}) and behaving asymptotically with the number of qubits in a similar fashion. They can therefore define a good alternative strategy to get highly entangled symmetric states.

\begin{figure}
\begin{center}
\includegraphics[width=.95\linewidth]{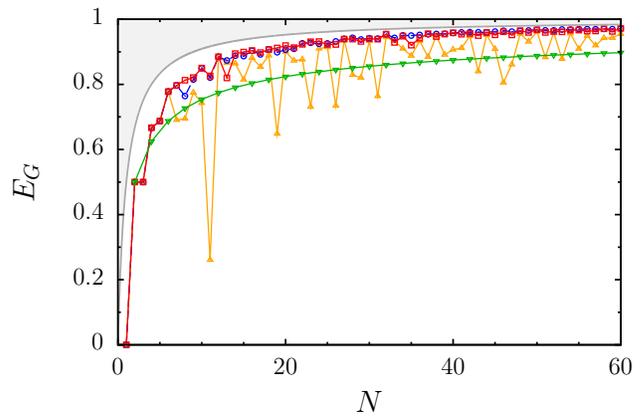}
\end{center}
\caption{Geometric entanglement of symmetric states for different arrangements~: Coulomb (blue circles), Tammes (orange triangles) and covering (red squares). The entanglement of Dicke states with a balanced number of excitations is displayed in green (reversed triangles). The grey shaded area shows the domain ruled out by the upper bound (\ref{upperbound}).}
\label{figTammes}
\end{figure}

We have investigated other configurations of Majorana representation points by considering other ways of distributing as evenly as possible $N$ points on a unit sphere. The Tammes and covering problems define such alternative ways. In the Tammes problem, one seeks to maximize the minimal distance $\min_{i\ne j}|\mathbf{r}_i-\mathbf{r}_j|$ among the $N$ points on the sphere. In the covering problem, one rather seeks to minimize $\max_{\mathbf{r}\in S}\min_i|\mathbf{r}-\mathbf{r}_i|$ that represents the greatest distance between a point $\mathbf{r}$ on the surface of a unit sphere $S$ and the nearest of the $N$ points. Tammes and covering arrangements are also given in Ref.~\cite{Sloane} for $N$ up to 130. For $N=2-6$ and $N=12$, Coulomb, Tammes and covering arrangements are identical~\cite{Lee57}. Figure~\ref{figTammes} shows the entanglement relative to these different arrangements together with the entanglement of symmetric Dicke states with a balanced number of excitations. Coulomb and covering arrangements give very similar results. In contrast, Tammes arrangement shows large fluctuations. The least fluctuating curve corresponds to Coulomb.

\subsection{Arrangements with equally weighted superpositions of Dicke states}

\begin{figure}
\begin{center}
\includegraphics[width=0.95\linewidth]{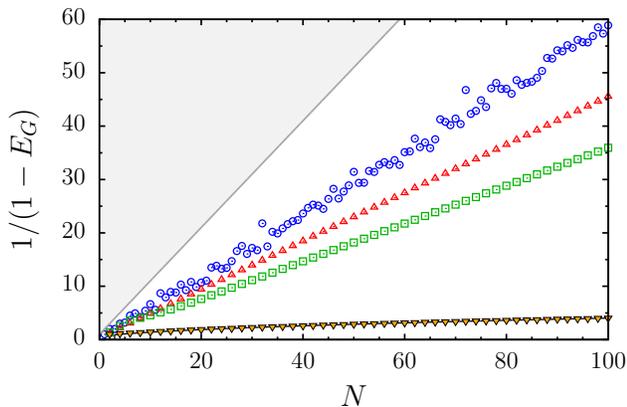}
\end{center}
\caption{Geometric entanglement for different sequences of symmetric states as a function of the number $N$ of qubits. Blue circle~: $E_G(\ket{\psi_{\mathrm{Coul}}(N)})$; red triangles~: $E_G(|\psi_{2/3}(N)\rangle)$; green squares~: $E_G(|\psi_{1}(N)\rangle)$; orange reverse triangles~: $E_G(|\psi_{2/3}^{\mathrm{lin}}(N)\rangle)$ fitted in excellent agreement (black curve) by Eq.~(\ref{eglin}). The grey shaded area shows the domain ruled out by the upper bound (\ref{upperbound}).}
\label{GEsuperposition}
\end{figure}

Equally weighted superpositions of Dicke states may exhibit interesting entanglement properties with respect to the geometric measure if their relative phases are properly chosen. We first consider the states
\begin{equation}
\label{psisymk2}
|\psi_\gamma(N)\rangle = \sum_{k=0}^N \frac{e^{i\gamma k^2}}{\sqrt{N+1}} |D_N(k) \rangle,
\end{equation}
with $\gamma$ a real parameter. The geometric entanglement of these states as a function of $N$ is displayed in Fig.~\ref{GEsuperposition} for two values of $\gamma$ ($2/3$ and $1$). It is compared to the geometric entanglement of the Coulomb arrangement. For the two values of $\gamma$ under consideration, the states (\ref{psisymk2}) are clearly less entangled than Coulomb. However, in contrast to the latter, the curves show interestingly almost no fluctuations and the entanglement behaves in a similar fashion for large $N$~: at least, for $N$ up to 100, we have
\begin{equation}\label{entpsisymk2}
E_G(\ket{\psi_\gamma(N)})\simeq 1-\frac{D_\gamma}{N + 1},
\end{equation}
with $D_\gamma$ a constant depending on $\gamma$ ($D_{2/3}\approx 2.22$, $D_1\approx 2.81$).

\begin{figure}
\begin{center}
\includegraphics[width=0.5\linewidth,bb = 97 700 155 755, clip=true]{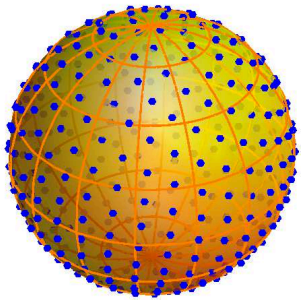}
\end{center}
\caption{Majorana representation of the state (\ref{psisymk2}) for $\gamma=2/3$ and $N=400$.}
\label{spiral}
\end{figure}

We obtained numerically that the Majorana representation points of the state (\ref{psisymk2}) spread in a kind of spiral on the Bloch sphere, with angles ($\theta_k,\varphi_k)$ given by
\begin{align}\label{tspiral}
&\theta_k\approx \arccos\left(1-\frac{2(k-1)}{N-1}\right),\\
&\varphi_k\approx \gamma(1-2k)~\mathrm{mod}~2\pi.\label{pspiral}
\end{align}
The corresponding Majorana representation is illustrated in Fig.~\ref{spiral} for $N=400$ and $\gamma = 2/3$. The large spread of the points explains the high geometric entanglement value obtained for that state.

We next consider another equally weighted superpositions of Dicke states with linearly increasing phases, namely
\begin{equation}
\label{psisymk1}
|\psi_{\gamma}^{\mathrm{lin}}(N)\rangle = \sum_{k=0}^N \frac{e^{i\gamma k}}{\sqrt{N+1}} |D_N(k) \rangle.
\end{equation}
These states have the advantage that their geometric entanglement can be analytically derived for large $N$.
Their overlap with the symmetric separable state $\ket{\Phi} \equiv \ket{\phi,\ldots, \phi}$ reads explicitly
\begin{equation}
|\bracket{\psi_{\gamma}^{\mathrm{lin}}(N)}{\Phi}|^2=\frac{1}{N+1}\left|\sum_{k=0}^N \sqrt{C^k_N\,p^{N-k}(1-p)^{k}}e^{i(k\varphi-\gamma k)}\right|^2\,.
\end{equation}
with $(\theta,\varphi)$ the two angles of the single qubit state $|\phi\rangle$.
The maximization over $\varphi$ yields straightforwardly
$\varphi=\gamma$. The remaining sum is well approximated by
replacing the binomial distribution with a Gaussian distribution, and
the sum by an integral.  The maximum of the Gaussian distribution is
situated at $N(1-p)$, and its variance is $Np(1-p)$, so as for sufficiently large $N$ the integration
range can be extended to $\pm\infty$. We then get
\begin{equation}
|\bracket{\psi_{\gamma}^{\mathrm{lin}}(N)}{\Phi}|^2\approx \frac{\sqrt{8\pi p(1-p)N}}{N+1},\label{eglinth}
\end{equation}
where $p=\cos^2\left(\theta/2\right)$. The overlap (\ref{eglinth}) is maximum for $\theta=\pi/2$ and yields the geometric entanglement
\begin{equation}
E_G(|\psi_{\gamma}^{\mathrm{lin}}(N)\rangle) \approx 1-\frac{\sqrt{2\pi N}}{N+1}\,.\label{eglin}
\end{equation}

Figure~\ref{GEsuperposition} shows that Eq.~(\ref{eglin}) reproduces the exact
numerical data very well, as soon as $N$ is of the order 10. Comparison with Eq.~(\ref{entDicke}) shows that superposing Dicke
states with linearly increasing phases reduces the geometrical
entanglement compared to a single Dicke state with $k$ equal (or close to) $N/2$.
The fact that a $1/\sqrt{N}$ behavior is obtained for linear phases,
but a $1/N$ behavior for quadratically increasing phases,
demonstrates that the phases of the Dicke states play a key role in
the amount of geometric entanglement that can be achieved.

\section{Geometric entanglement and quantumness}
\label{SecQuantumness}

In \cite{Q5} a quantumness measure was introduced in order to quantify how quantum an arbitrary mixed spin-$j$ quantum state $\rho$ is. The measure was defined as the Hilbert-Schmidt distance from the state $\rho$ to the set of classical states (which is the convex hull of spin coherent states). A spin-$j$ state can be seen as a symmetrized state of $2j$ spins $\frac{1}{2}$, and thus it can be written under the form Eq.~\eqref{psisym} with $N=2j$.  A spin-$j$ coherent state is the tensor product of $2j$ identical spin-$\frac{1}{2}$ coherent states. Since any spin-$\frac{1}{2}$ pure state is a coherent state,  spin-$j$ coherent states are by definition the symmetric separable pure states.

Let $\rho$ be a generic spin-$j$ density matrix. We define the B-quantumness $Q_B(\rho)$ as the distance from $\rho$ to the set $\mathcal{C}$ in the vector space of $(N+1)\times (N+1)$ matrices equipped with the Bures metric. Namely
\begin{equation}
\label{defqb}
Q_B(\rho)=\min_{\rho_C\in\mathcal{C}}D_B(\rho,\rho_C),
\end{equation}
where $D_B$ is the Bures distance~\cite{Bur69} defined by
\begin{equation}
D_B(\rho_A,\rho_B)=\sqrt{2-2\tr\sqrt{\sqrt{\rho_A}\rho_B\sqrt{\rho_A}}}.
\end{equation}
In other words, we consider a problem analogous to the one in~\cite{Q5} but
replace Hilbert-Schmidt distance by Bures distance.
If $\rho=\ket{\psi_S}\bra{\psi_S}$ is a pure state density matrix, with $\ket{\psi_S}$ a generic spin-$j$ state
defined as in Eq.~\eqref{psisym}, the expression for the  Bures distance simplifies drastically to
\begin{equation}
\label{dbpure}
D_B(\ket{\psi_S}\bra{\psi_S},\rho_C)=\sqrt{2-2\sqrt{\bra{\psi_S}\rho_C\ket{\psi_S}}}.
\end{equation}
Any element $\rho\in\mathcal{C}$ can be decomposed as a mixture of a certain number $n$ of coherent states $\ket{\Phi_i}$, i.e.
\begin{equation}
\rho_C=\sum_{i=1}^{n}\lambda_i\ket{\Phi_i}\bra{\Phi_i}
\end{equation}
with the $\lambda_i$ positive weights summing up to 1. The scalar product appearing in Eq.~\eqref{dbpure} then reads
\begin{equation}
\label{prp}
\bra{\psi_S}\rho_C\ket{\psi_S}=\sum_{i=1}^{n}\lambda_i|\langle\Phi_i|\psi_S\rangle|^2.
\end{equation}
B-quantumness is obtained by minimizing the quantity \eqref{dbpure} over $\mathcal{C}$, which is equivalent to maximizing the scalar product in \eqref{prp}. The maximum with respect to variation of the non-negative $\lambda_i$ occurs when all but one are equal to zero. The problem thus reduces to finding the largest overlap of the fixed $\ket{\psi_S}$ with a coherent state, and Eq.~\eqref{defqb} becomes
\begin{equation}
\label{qb-eg}
Q_B(\ket{\psi_S}\bra{\psi_S})=\sqrt{2-2\max_{\ket{\Phi}}|\langle\Phi|\psi_S\rangle|}.
\end{equation}
Thus B-quantumness and geometric entanglement \eqref{geoentsym} are
essentially the same quantity. The maximally entangled symmetric states
obtained in Sec.~\ref{SecHighGME} are thus the pure states with largest B-quantumness, which we call Queens of B-Quantumness (QBQ). As was mentioned
in~\cite{Q5} for Hilbert-Schmidt quantumness, one can easily
deduce from the fact that $Q_B$ is a convex function that B-quantumness
reaches its maximum for pure states, thus QBQ states can be looked for among
pure states. It is interesting to compare these QBQ to the most quantum
states (QQ states) obtained in \cite{Q5}.  For the lowest values of $N$, the
QBQ states differ from the QQ states only for $N=3$ and 5.

In the case of two qubits, entropy of entanglement is known to be the unique
entanglement measure for pure states~\cite{PopRoh97}. Therefore QBQ state for $N=2$ coincides with the
Dicke state $\ket{D_2(1)}$ which is also the QQ state (see Sec.~~\ref{SecHighGME} and Ref.~\cite{Q5}). For more than two qubits there is no unique measure of
entanglement. Thus one should not expect to find a unique measure of
quantumness, and the discrepancy between QBQ and QQ states should come as no surprise.

\section{Conclusion}
\label{SecConclusion}

In summary, we have realized a detailed study of the geometrical entanglement properties of symmetric $N$-qubit states. We have focused on the identification of symmetric states with a high geometric entanglement and how their entanglement behaves asymptotically for large $N$. We have shown that much higher geometric entanglement with improved asymptotical behavior can be obtained in comparison with the highly entangled balanced Dicke states~\cite{Wei03}. We have also derived an upper bound $E_G = 1-1/(N+1)$ for the geometric entanglement in the restricted set of the symmetric states. This value is significantly tighter than the one holding for any state. The connection with the quantumness of the states has been established.

\begin{acknowledgments}
This work has been supported by the Belgian
Institut Interuniversitaire des Sciences Nucl\'eaires (IISN). J.M.\
thanks the Belgian F.R.S.-FNRS for financial support. This work was supported in part by the Agence Nationale de la Recherche (ANR), project QPPRJCCQ. P.B. is grateful to the Sonderforschungsbereich TR 12 of the Deutsche Forschungsgemeinschaft and to the GDRI-471.
\end{acknowledgments}


\begin{thebibliography}{99}
\bibitem{Nie00} M. Nielsen and I. Chuang,  {\it Quantum Computation and Quantum Information} (Cambridge University Press, Cambridge, UK, 2000); R. Horodecki, P. Horodecki, M. Horodecki, and K. Horodecki, Rev. Mod. Phys. {\bf 81}, 865 (2009).
\bibitem{Ple07} M. B. Plenio and S. Virmani, Quantum Inf. Comput. {\bf 7}, 1 (2007); O. G\"uhne and G. T\'oth, Physics Reports {\bf 474}, 1 (2009).
\bibitem{Wei03} T.-C. Wei and P. M. Goldbart, Phys. Rev. A {\bf 68}, 042307 (2003); T.-C. Wei, M. Ericsson, P. M. Goldbart, and W. J. Munro,
Quantum Inf. Comput. {\bf 4}, 252 (2004).
\bibitem{Bih02} O. Biham, M. A. Nielsen, and T. Osborne, Phys. Rev. A {\bf 65}, 062312 (2002); Y. Shimony, D. Shapira, and O. Biham, Phys. Rev. A {\bf 69}, 062303 (2004).
\bibitem{Hay06} M. Hayashi, D. Markham, M. Murao, M. Owari, and S. Virmani, Phys. Rev. Lett. {\bf 96}, 040501 (2006).
\bibitem{Hay08} M. Hayashi, D. Markham, M. Murao, M. Owari, and S. Virmani, Phys. Rev. A {\bf 77}, 012104 (2008).
\bibitem{Wei05} T.-C. Wei, D. Das, S. Mukhopadyay, S. Vishveshwara, and P. M. Goldbart, Phys. Rev. A {\bf 71}, 060305 (2005).
\bibitem{Mar08} D. Markham, J. Anders, V. Vedral, M. Murao, A. Miyake, Euro. Phys. Lett. {\bf 81}, 40006 (2008).
\bibitem{Oru08} R. Or\'us, S. Dusuel, and J. Vidal, Phys. Rev. Lett. {\bf 101}, 025701 (2008).
\bibitem{Nak09} Y. Nakata, D. Markham, M. Murao, Phys. Rev. A {\bf 79}, 042313 (2009).
\bibitem{Tam10} S. Tamaryan, A. Sudbery, and L. Tamaryan, arXiv:1002.3049.
\bibitem{Mar07} D. Markham, A. Miyake, S. Virmani, New J. Phys. {\bf 9}, 194 (2007); X. Y. Chen, arXiv:0909.1603.
\bibitem{Tam08} L. Tamaryan, D. K. Park, and S. Tamaryan, Phys. Rev. A {\bf 77}, 022325 (2008); 
S. Tamaryan, T.-C. Wei, and D. Park, Phys. Rev. A {\bf 80}, 052315 (2009); 
S. Tamaryan, H. Kim, M. S. Kim, K. S. Jang, and D. K. Park, arXiv:0909.1077; 
\bibitem{Mar09} M. Hayashi, D. Markham, M. Murao, M. Owari, and S. Virmani, J. Math. Phys. {\bf 50}, 122104 (2009).
\bibitem{Par09} P. Parashar and S. Rana, arXiv:0909.4443.
\bibitem{Che09} L. Chen, A. Xu, and H. Zhu, arXiv:0911.1493.
\bibitem{Hub09} R. H{\"u}bener, M. Kleinmann, T.-C. Wei, C. Gonz{\'a}lez-Guill{\'e}n and O. G{\"u}hne, Phys. Rev. A {\bf 80}, 032324 (2009).
\bibitem{Jun08} E. Jung, M.-R. Hwang, H. Kim, M.-S. Kim, D. Park, J.-W. Son, and S. Tamaryan, Phys. Rev. A {\bf 77}, 062317 (2008).
\bibitem{Q5} O. Giraud, P. Braun, and D. Braun, arXiv:1002.2158 

\bibitem{Maj32} E. Majorana, Nuovo Cimento {\bf 9}, 43 (1932).
\bibitem{Dic54} R. H. Dicke, Phys. Rev. {\bf 93}, 99 (1954).
\bibitem{Ben06} see, e.g., I. Bengtsson and K. Zyczkowski, Geometry of Quantum States: An Introduction to Quantum Entanglement (2006).
\bibitem{Bas09a} T. Bastin, C. Thiel, J. von Zanthier, L. Lamata, E. Solano, and G. S. Agarwal, Phys. Rev. Lett. {\bf 102}, 053601 (2009).
\bibitem{Bas09b} T. Bastin, S. Krins, P. Mathonet, M. Godefroid, L. Lamata, and E. Solano, Phys. Rev. Lett. {\bf 103}, 070503 (2009).
\bibitem{Bar97} S. M. Barnett and P. M. Radmore, Methods in Theoretical Quantum Optics, Oxford University Press, Oxford (1997).
\bibitem{Mar10} On completion of this work, we became aware of a similar study in progress, Ref.~[11] of D. Markham, arXiv:1001.0343.
\bibitem{Sloane} N. J. A. Sloane, R. H. Hardin, W. D. Smith and others, Tables of Spherical Codes, published electronically at www.research.att.com/{$\sim$}njas/packings/.
\bibitem{Lee57} J. Leech, Math. Gazette {\bf 41}, 81 (1957).

\bibitem{Bur69} D. Bures, Trans. Am. Math. Soc. {\bf 135}, 199 (1969).
\bibitem{PopRoh97} S. Popescu and D. Rohrlich, Phys. Rev. A {\bf 56}, R3319 (1997).

\end{thebibliography}
\end{document}